\documentclass[letterpaper,twocolumn,10pt]{article}
\usepackage{usenix2019_v3}

\usepackage{tikz}
\usepackage{amsmath}

\usepackage{filecontents}

\usepackage{booktabs} 
\usepackage{color}
\usepackage{amsmath}
\usepackage{algorithm}
\usepackage[noend]{algpseudocode}

\usepackage[inline]{enumitem}
\usepackage{listings}
\usepackage{subcaption}
\usepackage{balance}
\usepackage{comment}
\usepackage{graphicx}
\usepackage{tikz}

\definecolor{mGreen}{rgb}{0,0.6,0}
\definecolor{mGray}{rgb}{0.5,0.5,0.5}
\definecolor{mPurple}{rgb}{0.58,0,0.82}
\definecolor{backgroundColour}{rgb}{0.99,0.99,0.99}

\lstdefinestyle{CStyle}{
	backgroundcolor=\color{backgroundColour},   
	commentstyle=\color{mGreen},
	keywordstyle=\color{magenta},
	numberstyle=\tiny\color{mGray},
	stringstyle=\color{mPurple},
	basicstyle=\footnotesize,
	breakatwhitespace=false,         
	breaklines=true,                 
	captionpos=b,                    
	keepspaces=true,                 
	numbers=left,                    
	numbersep=5pt,                  
	showspaces=false,                
	showstringspaces=false,
	showtabs=false,                  
	tabsize=2,
	xleftmargin=2em,
	framexleftmargin=1.5em,
	language=C
}

\usepackage{pdfpages}
\usetikzlibrary{calc,positioning,arrows}

\newcommand{\paperoutline}[1]{\textcolor{orange}{#1}}

\newcommand{\ie}{\textit{i.e.,} }
\newcommand{\eg}{\textit{e.g.,} }

\newcommand{\todobox}[3]{%
	\colorbox{#1}{\textcolor{white}{\sffamily\bfseries\scriptsize #2}}%
	~\textcolor{blue}{#3} %
	\textcolor{#1}{$\triangleleft$}%
}
\newcommand{\todo}[1]{\todobox{red}{TODO}{#1}}

\begin{document}

\date{}

\title{\Large \bf ScaRR: \\ Scalable Runtime Remote Attestation for Complex Systems}

\author{
 {\rm Flavio Toffalini}\\
 {SUTD}
  \and
 {\rm Eleonora Losiouk}\\
 {University of Padua}
 \and
 {\rm Andrea Biondo}\\
 {University of Padua}
 \and
 {\rm Jianying Zhou}\\
 {SUTD}
 \and
 {\rm Mauro Conti}\\
 {University of Padua} \\ \\
 {\rm flavio\_toffalini@mymail.sutd.edu.sg, jianyingzhou@sutd.edu.sg}\\
 {\rm elosiouk@math.unipd.it, andrea.biondo.1@studenti.unipd.it, conti@unipd.it} \\
} 

\maketitle
\begin{abstract}
The introduction of remote attestation (RA) schemes has allowed academia and industry to enhance the security of their systems.
The commercial products currently available enable only the validation of static properties,
such as applications fingerprint, and do not handle runtime properties,
such as control-flow correctness.
This limitation pushed researchers towards the identification of new approaches, called runtime RA.
However, those mainly work on embedded devices,
which share very few common features with complex systems, such as virtual machines in a cloud. 
A naive deployment of runtime RA schemes for embedded devices on complex systems faces scalability problems,
such as the representation of complex control-flows or slow verification phase.  

In this work, we present ScaRR: the first Scalable Runtime Remote attestation schema for complex systems. Thanks to its novel control-flow model, ScaRR enables the deployment of runtime RA on any application regardless of its complexity, by also achieving good performance. We implemented ScaRR and tested it on the benchmark suite SPEC CPU 2017. 
We show that ScaRR can validate on average $2$M control-flow events per second, definitely outperforming existing solutions that support runtime RA on complex systems. 

\end{abstract}

\section{Introduction}
\label{sec:introduction}


RA is a procedure that allows an entity (\ie the \emph{Verifier}) 
to verify the status of a device (\ie the \emph{Prover}) from a remote location. This is achieved by having first the \emph{Verifier} sending a challenge to the \emph{Prover}, which replies with a report. 
Then, the \emph{Verifier} analyzes the report to identify whether the \emph{Prover} has been compromised~\cite{anati2013innovative}. 
In standard RA, usually defined as static, the \emph{Prover} verification involves the integrity of specific hardware and software properties (\eg the \emph{Prover} has loaded the correct software).
On the market, there are already several available products implementing static RA, such as Software Guard Extensions (SGX)~\cite{costan2016intel} or Trusted Platform Module (TPM)~\cite{tomlinson2017introduction}.
However, these do not provide a defence against runtime attacks (\eg the control-flow ones)
that aim to modify the program runtime behaviour. 
Therefore, to identify \emph{Prover} runtime modifications, researchers proposed runtime RA. Among the different solutions belonging to this category, there are also the control-flow attestation approaches, which
encode the information about the executed control-flow of a process~\cite{abera2016c,aberadiat}.

In comparison to static RA, the runtime one is relatively new, and today there are no reliable products available on the market since researchers have mainly investigated runtime RA for embedded devices~\cite{abera2016c,zeitouni2017atrium,aberadiat,dessouky2017fat,Dessouky:2018:LLH:3240765.3240821}: 
most of them encode the complete execution path of a \emph{Prover} in a single hash~\cite{abera2016c,zeitouni2017atrium,dessouky2017fat}; 
some~\cite{aberadiat} compress it in a simpler representation and rely on a policy-based verification schema; 
other ones~\cite{Dessouky:2018:LLH:3240765.3240821} adopt symbolic execution to verify the control-flow information continuously sent by the \emph{Prover}. 
Even if they have different performances, none of the previous solutions can be applied to a complex system (\eg virtual machines in a cloud) due to the following reasons: 
\begin{enumerate*}[label=(\roman*)]
	\item representing all the valid execution paths through hash values is unfeasible (\eg the number of execution paths tends to grow exponentially with the size of the program),
	\item the policy-based approaches might not cover all the possible attacks,
	\item symbolic execution slows down the verification phase.
\end{enumerate*}


The purpose of our work is to fill this gap by providing ScaRR, the first runtime RA schema for complex systems.
In particular, we focus on environments such as Amazon Web Services~\cite{aws} or Microsoft Azure~\cite{azure}. 
Since we target such systems, we require support for features such as multi-threading.
Thus, ScaRR provides the following achievements with respect over the current solutions supporting runtime RA: 
\begin{enumerate*}[label=(\roman*)]
	\item it makes the runtime RA feasible for any software,
	\item it enables the \emph{Verifier} to verify intermediate states of the \emph{Prover} without interrupting its execution
	\item it supports a more fine-grained analysis of the execution path where the attack has been performed. 
\end{enumerate*}
We achieve these goals thanks to a novel model for representing the execution paths of a program, which is based on the fragmentation of the whole path into meaningful sub-paths. As a consequence, the \emph{Prover} can send a series of intermediate partial reports, which are immediately validated by the \emph{Verifier} thanks to the lightweight verification procedures performed.  



ScaRR is designed to defend a \emph{Prover}, equipped with a trusted anchor and with a set of the standard solutions (\eg W$\oplus$X/DEP~\cite{pinzari2003introduction},
Address Space Layout Randomization (ASLR)~\cite{kil2006address} and Stack Canaries~\cite{baratloo2000transparent}), from attacks performed in the user-space and aimed at modifying the \emph{Prover} runtime behaviour. The current implementation of ScaRR requires the program source code to be properly instrumented through a compiler based on LLVM~\cite{lattner2004llvm}. However, it is possible to use lifting techniques~\cite{mcsema}, as well. 
Once deployed, ScaRR allows to verify on average $2M$ control-flow events per second, which is significantly more than the few hundred 
per second~\cite{Dessouky:2018:LLH:3240765.3240821} or the thousands per second~\cite{aberadiat} verifiable through the existing solutions.

\textbf{Contribution.} The contributions of this work are the following ones: 
\begin{itemize}
	\item We designed a new model for representing the execution path for applications of any complexity.
	\item We designed and developed ScaRR, the first schema that supports runtime RA for complex systems.
	\item We evaluated the ScaRR performances in terms of: 
	\begin{enumerate*}[label=(\roman*)]
		\item attestation speed (\ie the time required by the \emph{Prover} to generate a partial report),
		\item verification speed (\ie the time required by the \emph{Verifier} to evaluate a partial report),
		\item overall generated network traffic (\ie the network traffic generated during the communication between \emph{Prover} and \emph{Verifier}).
	\end{enumerate*}
\end{itemize}

\vspace{3mm}
\textbf{Organization.}
The paper is organized as follow.
First, we provide a background on standard RA and control-flow exploitation (Section~\ref{sec:background}), and define the threat model (Section~\ref{sec:threat-model}).
Then, we describe the ScaRR control-flow model (Section~\ref{sec:model}) and its design (Section~\ref{sec:proposal}).
We discuss ScaRR implementation details (Section~\ref{sec:implementation})
and evaluate its performance and security guarantees (Section~\ref{sec:evaluation}).
Finally, we discuss ScaRR limitations (Section~\ref{sec:discussion}), related works (Section~\ref{sec:related-works}), and conclude with final remarks (Section~\ref{sec:conclusion}).
\section{Background}
\label{sec:background}

The purpose of this section is to provide background knowledge about standard RA procedures and control-flow attacks. 

\paragraph{Remote Attestation.}
RA always involves a \emph{Prover} and a \emph{Verifier}, with the latter responsible for verifying the current status of the former. Usually, the \emph{Verifier} sends a challenge to the \emph{Prover} asking to measure specific properties. The \emph{Prover}, then, calculates the required measurement (\eg a hash of the application loaded) and sends back a report R, which contains the measurement M along with a digital fingerprint F, for instance, $R = (M,F)$. Finally, the \emph{Verifier} evaluates the report, considering its freshness (\ie the report has not been generated through a replay attack) and correctness (\ie the \emph{Prover} measurement is valid). It is a standard assumption that the \emph{Verifier} is trusted, while the \emph{Prover} 
might be compromised. However, the \emph{Prover} is able to generate a correct and fresh report due to its trusted anchor (\eg a dedicated hardware module).

\paragraph{Control-Flow Attacks.}

\begin{figure}[b]
\centering
\begin{subfigure}[t]{0.25\textwidth}
	\centering
	\includegraphics[width=\linewidth]{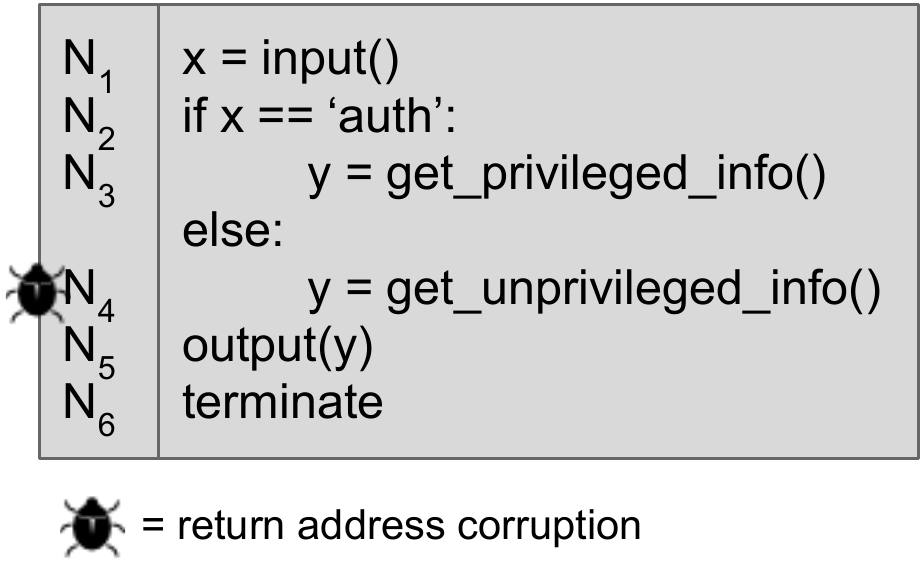}
	\caption{Pseudo-code of a program under a control-flow attack.}
	\label{fig:problem-setting-code}
\end{subfigure}
~
\begin{subfigure}[t]{0.2\textwidth}
	\centering
	\includegraphics[width=\linewidth]{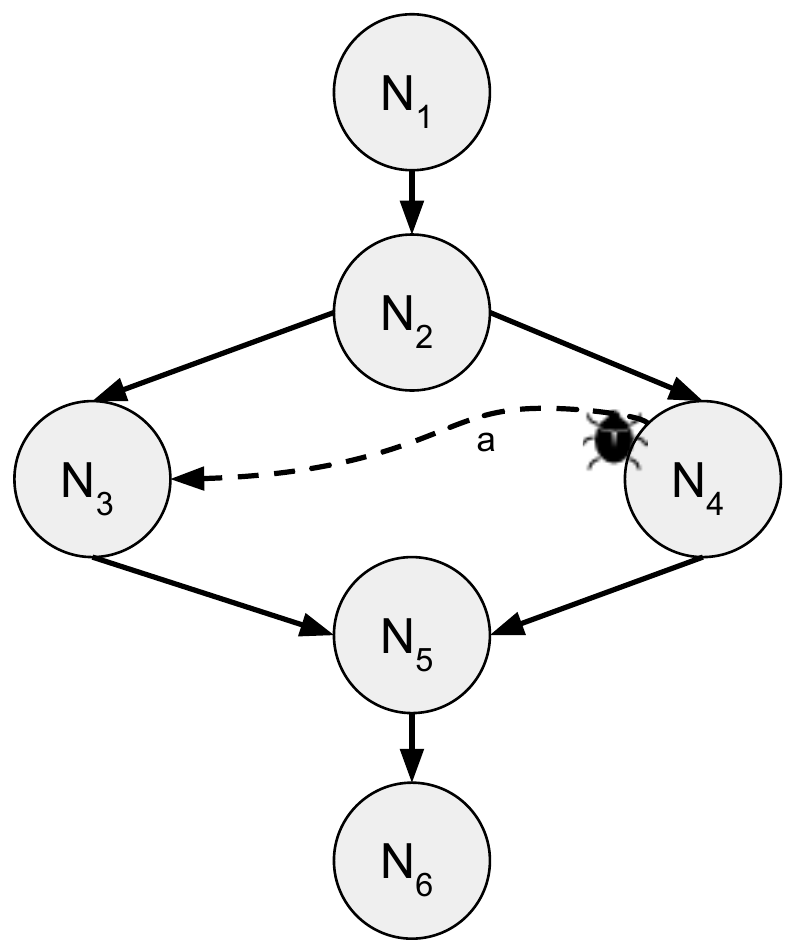}
	\caption{Control-flow graph of a program under a control-flow attack.}
	\label{fig:problem-setting-graph}
\end{subfigure}
\caption{Illustrative example of a control-flow attack.}
\label{fig:problem-setting}
\end{figure}

To introduce control-flow attacks, we first discuss the concepts of control-flow graph~(CFG), execution-path, and basic-block~(BBL) by using the simple program shown in Figure~\ref{fig:problem-setting-code} as a reference example. 
The program starts with the acquisition of an input from the user (line $1$). This is evaluated (line $2$) in order to redirect the execution towards the retrieval of a privileged information (line $3$) or an unprivileged one (line $4$). Then, the retrieved information is stored in a variable ($y$), which is returned as an output (line $5$), before the program properly concludes its execution (line $6$). 

A CFG represents all the paths that a program
may traverse during its execution and it is statically computed. 
On the contrary, an execution path is a single path of the CFG traversed by the program at runtime. 
The CFG associated to the program in Figure~\ref{fig:problem-setting-code} is depicted in
Figure~\ref{fig:problem-setting-graph} and it encompasses two components: nodes and edges. The former are the BBLs of the program, while the latter represent the standard flow traversed by the program to move from a BBL towards the next one. A BBL is a linear sequence of instructions with a single entry point
(\ie no incoming branches to the set of instructions other than the first), 
and a single exit point (\ie no outgoing branches from the set of instructions other than the last). Therefore, a BBL can be considered an atomic unit with respect to the control-flow, as it will either be fully executed, or not executed at all on a given execution path.
A BBL might end with a control-flow event, which could be one of the following in a \texttt{X$86\_64$} architecture: procedure calls (\eg \texttt{call}), jumps (\eg \texttt{jmp}), procedure returns (\eg \texttt{ret}), and system calls (\eg \texttt{syscall}). 
During its execution, a process traverses several BBLs, which completely define the process execution path.

Runtime attacks, and more specifically the control-flow ones, aim at modifying the CFG of a program by tampering with its execution path.
Considering Figure~\ref{fig:problem-setting}, we assume that an attacker is able to run the program (from the node $N_1$), but that he is not authorized to retrieve the privileged information.
However, the attacker can, anyway, violate those controls through a memory corruption error performed on the node $N_4$.
As soon as the attacker provides an input to the program and starts its execution, he will be redirected to the node $N_4$.
At this point, the attacker can exploit a memory corruption error (\eg a stack overflow) to introduce a new edge from $N_4$ to $N_3$ (edge labeled as $a$) and retrieve the privileged information.
As a result, the program traverses an unexpected execution path not belonging to its original CFG.
Even though several solutions have been proposed to mitigate such attacks (\eg ASLR~\cite{kil2006address}), attackers still manage to perform them~\cite{van2012memory}. 


This illustrative example about how to manipulate the execution path of a program is usually the basic step to perform more sophisticated attacks like
exploiting a vulnerability to take control of a process~\cite{yuan2015hardware} or installing a persistent data-only malware without injecting new code, 
once the control over a process is taken by the attacker~\cite{vogl2014persistent}.

Runtime RA provide a reliable mechanism which allows the \emph{Verifier} to trace and validate the execution path undertaken by the \emph{Prover}. 
\section{Threat Model and Requirements}
\label{sec:threat-model}


In this section, we describe the features of the \emph{Attacker} and the \emph{Prover} involved in out threat model. Our assumptions are in line with other RA schemes~\cite{costan2016intel,winter2008trusted,abera2016c,aberadiat,Dessouky:2018:LLH:3240765.3240821}.

\paragraph{Attacker.}
We assume to have an attacker that aims to control a remote service, such as a Web Server or a DBMS, and that has already bypassed the default protections, such as Control Flow Integrity (CFI). To achieve his aim, the attacker can adopt different techniques, among which: Return-Oriented Programming (ROP)/ Jump-Oriented Programming (JOP) attacks~\cite{carlini2014rop,bletsch2011jump}, function hooks~\cite{7778160}, injection of a malware into the victim process, installation of a data-only malware in user-space~\cite{vogl2014persistent}, or manipulation of other user-space processes, such as security monitors. 
In our threat model, we do not consider physical attacks (our complex systems
are supposed to be virtual machines), pure data-oriented attacks (\eg attacks
that do not alter the original program CFG), self-modifying code, and dynamic loading of code at runtime (\eg just-in-time compilers~\cite{suganuma2000overview}).

\paragraph{Prover.}
The \emph{Prover} is assumed to be equipped with:
\begin{enumerate*}[label=(\roman*)]
	\item a trusted anchor that guarantees a static RA,
	\item standard defence mitigation techniques, such as W$\oplus$X/DEP, ASLR.
\end{enumerate*}
In our implementation, we use the kernel as a trusted anchor, which is a reasonable assumption if the machines have trusted modules such as a TPM~\cite{tomlinson2017introduction}. 
However, we can also use a dedicated hardware, as discussed in Section~\ref{sec:discussion}. The \emph{Prover} maintains sensitive information (\ie shared keys and cryptographic functions) in the trusted anchor and uses it to generate fresh reports, that cannot be tampered by the attacker. 



\section{ScaRR Control-Flow Model}
\label{sec:model}

ScaRR is the first schema that allows to apply runtime RA on complex systems. To achieve this goal, it relies on a new model for representing the CFG/execution path of a program. In this section, we illustrate first the main components of our control-flow model (Section~\ref{ssec:basic_concepts}) and, then, the challenges we faced during its design (Section~\ref{ssec:challenges}).

%
%
%


\subsection{Basic Concepts}
\label{ssec:basic_concepts}

The ScaRR control-flow model handles BBLs at assembly level and involves two components: \emph{checkpoints} and \emph{List of Actions (LoA)}.

A \emph{checkpoint} is a special BBL used as a delimiter for identifying the start or the end of a sub-path within the CGF/execution path of a program. A \emph{checkpoint} can be: \emph{thread beginning/end}, if it identifies the beginning/end of a thread; \emph{exit-point}, if it represents an exit-point from an application module (\eg a system call or a library function invocation); \emph{virtual-checkpoint}, if it is used for managing special cases such as \emph{loops} and \emph{recursions}. 

A \emph{LoA} is the series of significant edges that a process traverses to move from a \emph{checkpoint} to the next one. Each edge is represented through its source and destination BBL and, comprehensively, a \emph{LoA} is defined through the following notation:
$$
[(\text{BBL}_{s1},\text{BBL}_{d1}), \dots, (\text{BBL}_{sn},\text{BBL}_{dn})].
$$
Among all the edges involved in the complete representation of a CFG, we consider only a subset of them.
In particular, we look only at those edges that identify a unique execution path: procedure call, procedure return and branch (\ie conditional and indirect jumps). 


To better illustrate the ScaRR control-flow model, we now recall the example introduced in Section~\ref{sec:background}. 
Among the six nodes belonging to the CFG of the example, only the following four ones are \emph{checkpoints}: $N_1$, since it is a \emph{thread beginning}; $N_3$ and $N_4$, because they are \emph{exit-points}, and $N_6$, since it is a \emph{thread end}. In addition, the \emph{LoAs} associated to the example are the following ones:  
\begin{equation*}
\begin{split}
N_1-N_3 &\Rightarrow [(N_2, N_3)] \\    
N_1-N_4 &\Rightarrow [(N_2, N_4)] \\ 
N_3-N_6 &\Rightarrow [] \\
N_4-N_6 &\Rightarrow [].
\end{split}
\end{equation*}
On the left we indicate a pair of \emph{checkpoints} (\eg $N_1-N_3$), while on the right the associated \emph{LoA} (empty \emph{LoAs} are considered valid).



\subsection{Challenges}
\label{ssec:challenges}
\emph{Loops}, \emph{recursions}, \emph{signals}, and \emph{exceptions} involved in the execution of a program introduce new challenges in the representation of a CFG since they can generate uncountable executions paths. For example, \emph{loops} and \emph{recursions} can generate an indefinite number of possible combinations of \emph{LoA}, while \emph{signals}, as well as \emph{exceptions}, can introduce an unpredictable execution path at any time. 

\begin{figure}[b]
	\centering
	\includegraphics[width=0.25\textwidth]{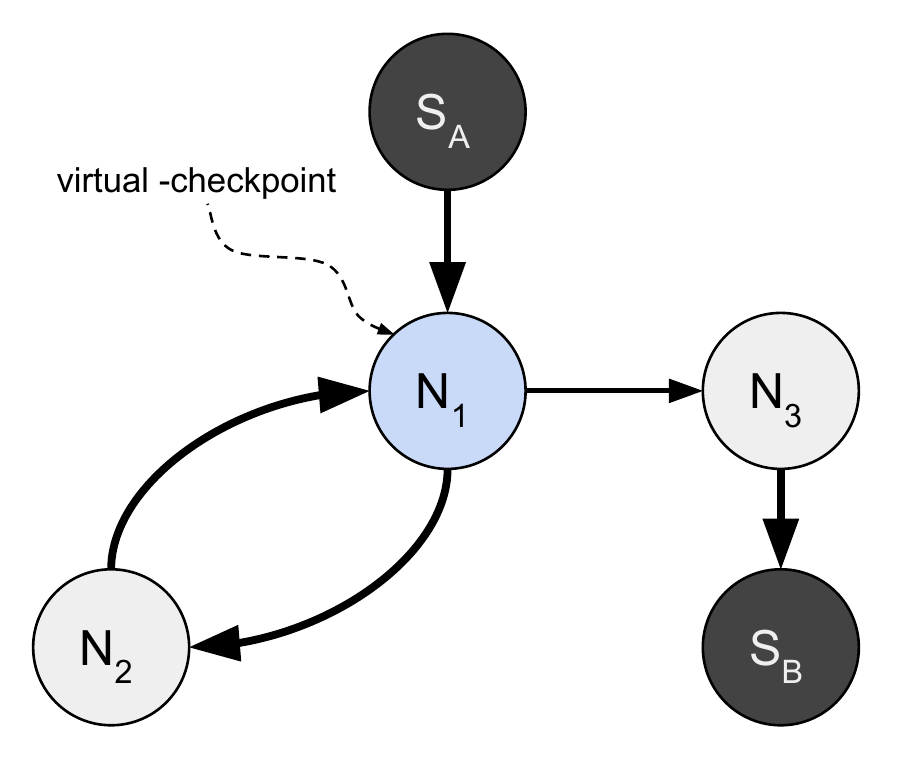}
	\caption{Loop example in the ScaRR control-flow model.}
	\label{fig:challenge-III}
\end{figure}

\textbf{Loops.}
In Figure~\ref{fig:challenge-III}, we illustrate the approach used to handle \emph{loops}.
Since it is not always possible to count the number of iterations of a loop, we consider the conditional node of the \emph{loop} (\texttt{$N_1$}) as a \emph{virtual-checkpoint}. 
Thus, the \emph{LoAs} associated to the example shown in Figure~\ref{fig:challenge-III} are as follows: 
\begin{equation*}
\begin{split}
S_A-N_1 &\Rightarrow [] \\    
N_1-N_1 &\Rightarrow [(N_1, N_2)] \\ 
N_1-S_B &\Rightarrow [(N_1, N_3)].
\end{split}
\end{equation*}

\begin{figure}[t]
	\centering
	\includegraphics[width=0.4\textwidth]{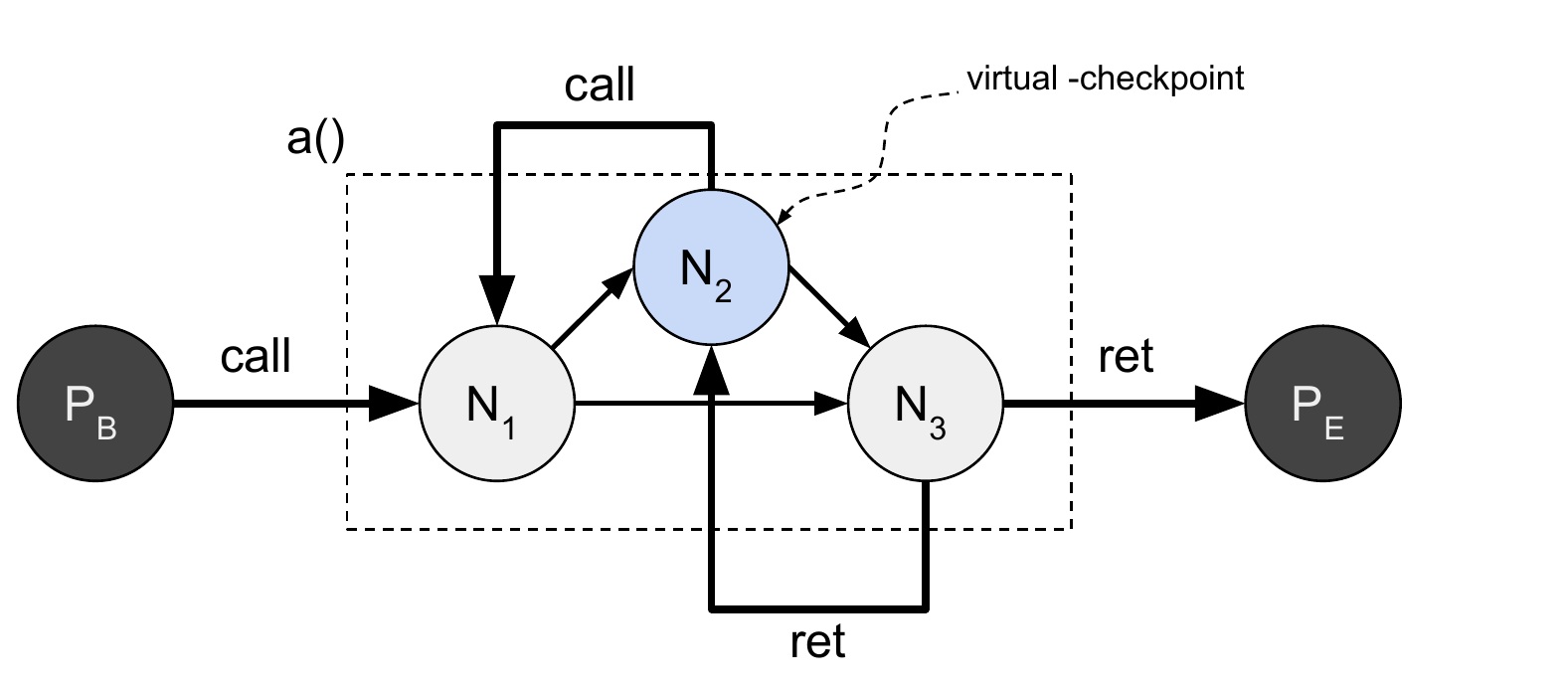}
	\caption{Recursion example in the ScaRR control-flow model.}
	\label{fig:challenge-IV}
\end{figure}

\textbf{Recursions.}
In Figure~\ref{fig:challenge-IV}, we illustrate our approach to handle \emph{recursions},
\ie a function that invokes itself. 
Intuitively, the \emph{LoAs} connecting \texttt{$P_B$} and \texttt{$P_E$} should contain all the possible invocations made by \texttt{a()} towards itself, but the number of invocations is indefinite. Thus, we consider the node performing the recursion as a \emph{virtual-checkpoint} and model only the path that could be chosen, without referring to the number of times it is really undertaken. The resulting \emph{LoAs} for the example in Figuree~\ref{fig:challenge-IV} as the following ones: 
\begin{equation*}
\begin{split}
P_B-N_2 &\Rightarrow [(P_B, N_1),(N_1,N_2)] \\    
N_2-N_2 &\Rightarrow [(N_2, N_1),(N_1,N_2)] \\ 
N_2-N_2 &\Rightarrow [(N_2, N_1),(N_1, N_3),(N_3, N_2)] \\
N_2-P_E &\Rightarrow [(N_2, N_1),(N_1, N_3),(N_3, P_E)] \\
P_B-P_E &\Rightarrow [(P_B, N_1),(N_1, N_3),(N_3, P_E)].
\end{split}
\end{equation*}

Finally, the \emph{virtual-checkpoint} can be used as a general approach to solve every situation in which an indirect jump targets a node already present in the \emph{LoA}.

\textbf{Signals.}
When a thread receives a signal, its execution is stopped and, after a context-switch, it is diverted to a dedicated handler (\eg a function).
This scenario makes the control-flow unpredictable, since an interruption can occur at any point during the execution. To manage this case, ScaRR models the signal handler as a separate thread (adding \emph{beginning/end thread checkpoints})
and computes the relative \emph{LoAs}. If no handler is available for the signal that interrupted the program, the entire process ends immediately, producing a wrong \emph{LoA}. 


\textbf{Exception Handler.}
Similar to signals, when a thread rises an exception, the execution path is stopped
and control is transferred to a catch block. 
Since ScaRR has been implemented for Linux,
we model the catch blocks as a separate thread (adding \emph{beginning/end thread checkpoints}),
but it is also possible to adapt ScaRR to fulfill different exception handling mechanisms (\eg in Windows).
In case no catch block is suitable for the exception that was thrown, the process gets interrupted and the generated \emph{LoA} is wrong.

\section{System Design}
\label{sec:proposal}

To apply runtime RA on a complex system, there are two fundamental requirements: 
\begin{enumerate*}[label=(\roman*)]
	\item handling the representation of a complex CFG or execution path,
	\item having a fast verification process.
\end{enumerate*}
Previous works have tried to achieve the first requirement through different approaches. A first solution~\cite{abera2016c,zeitouni2017atrium,dessouky2017fat} is based on the association of all the valid execution paths of the \emph{Prover} with a single hash value. 
Intuitively, this is not a scalable approach because it does not allow to handle complex CFG/execution paths. 
On the contrary, a second approach~\cite{Dessouky:2018:LLH:3240765.3240821} relies on the transmission of all the control-flow events to the \emph{Verifier}, 
which then applies a symbolic execution to validate their correctness. While addressing the first requirement, this solution suffers from a slow verification phase, which leads toward a failure in satisfying the second requirement. 

Thanks to its novel control-flow model, ScaRR enables runtime RA for complex systems, since its design specifically considers the above-mentioned requirements with the purpose of addressing both of them. In this section, we provide an overview of the ScaRR schema (Section~\ref{ssec:scarr_overview}) together with the details of its workflow (Section~\ref{ssec:scarr_details}), explicitly motivating how we address both the requirements needed to apply runtime RA on complex systems. 

\subsection{Overview}
\label{ssec:scarr_overview}
\begin{figure*}[h]
	\centering
	\includegraphics[width=0.6\textwidth]{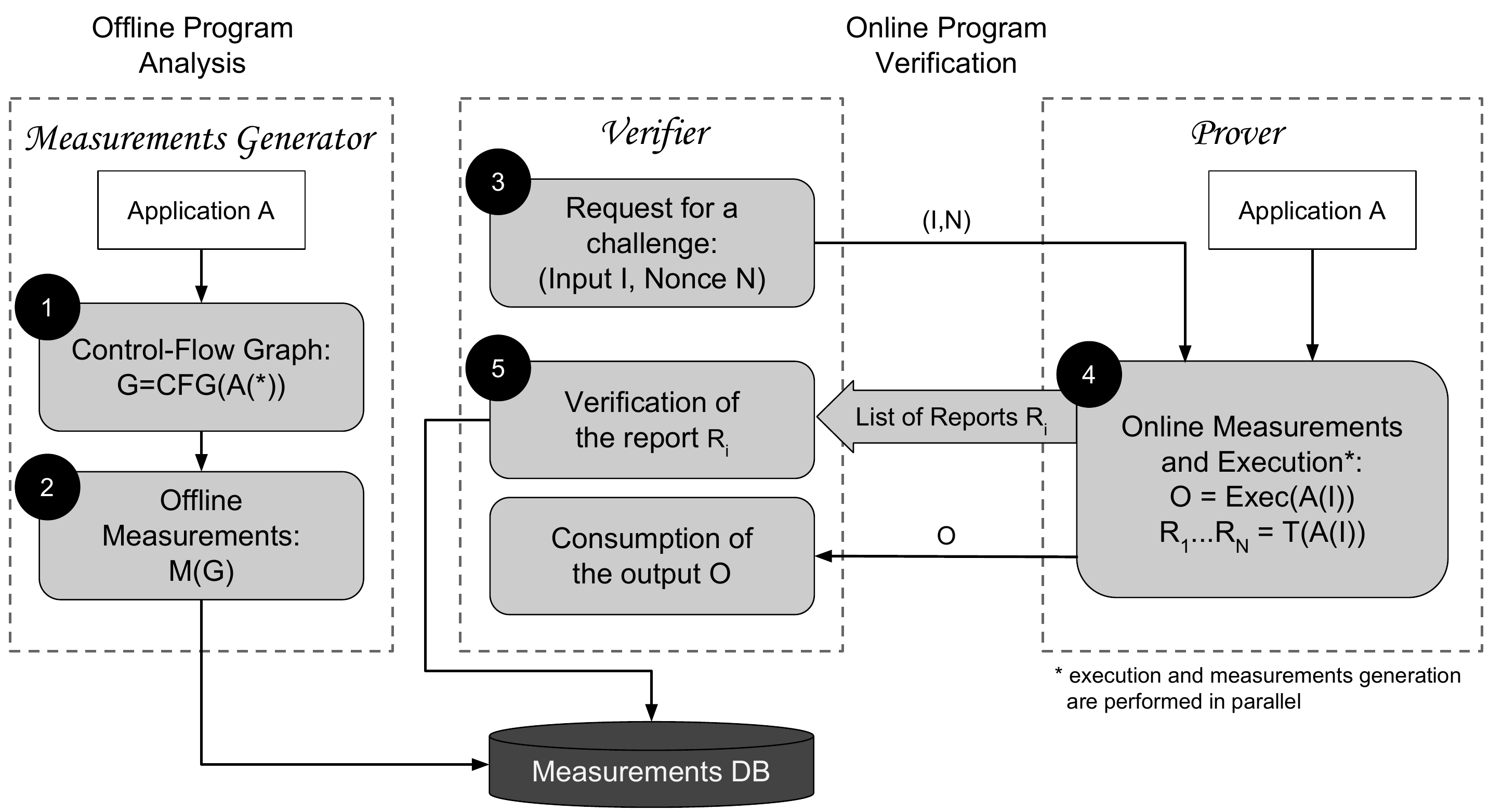}
	\caption{ScaRR system overview.}
	\label{fig:overview}
\end{figure*}

Even if the ScaRR control-flow model is composed of \emph{checkpoints} and \emph{LoAs}, the ScaRR schema relies on a different type of elements, which are the \emph{measurements}. Those are a combination of \emph{checkpoints} and \emph{LoAs} and contain the necessary information to perform runtime RA. 
Figure~\ref{fig:overview} shows an overview of ScaRR, which encompasses the following four components: a \emph{Measurements Generator}, for identifying all the program valid \emph{measurements}; 
a \emph{Measurements DB}, for saving all the program valid \emph{measurements};
a \emph{Prover}, which is the machine running the monitored program;
a \emph{Verifier}, which is the machine performing the program runtime verification. 

As a whole, the workflow of ScaRR involves two separate phases: an \emph{Offline Program Analysis} and an \emph{Online Program Verification}. 
During the first phase, the \emph{Measurements Generator} calculates the CFG of the monitored \emph{Application A} (Step 1 in Figure~\ref{fig:overview}) and, after generating all the \emph{Application A} valid \emph{measurements}, it saves them in the \emph{Measurements DB} (Step 2 in Figure~\ref{fig:overview}).
During the second phase, the \emph{Verifier} sends a challenge to the \emph{Prover} (Step 3 in Figure~\ref{fig:overview}).
Thus, the \emph{Prover} starts executing the  \emph{Application A} and sending partial reports to the \emph{Verifier} (Step 4 in Figure~\ref{fig:overview}). The \emph{Verifier} validates the freshness and correctness of the partial reports by comparing the received new \emph{measurements} with the previous ones stored in the \emph{Measurements DB}. Finally, as soon as the \emph{Prover} finishes the processing of the input received from the \emph{Verifier}, it sends back the associated output. 



\subsection{Details}
\label{ssec:scarr_details}

As shown in Figure~\ref{fig:overview}, the workflow of ScaRR goes through five different steps.
Here, we provide details for each of those.

\textbf{(1) Application CFG.} 
The \emph{Measurements Generator} executes the \emph{Application A()}, or a subset of it (\eg a function), and extracts the associated CFG $G$. 

\textbf{(2) Offline Measurements.} 
After generating the CFG, the \emph{Measurements Generator} computes all the program \emph{offline measurements} during the \emph{Offline Program Analysis}. Each \emph{offline measurement} is represented as a key-value pair as follows: 
$$
(\text{cp}_A,\text{cp}_B,H(\text{LoA})) \Rightarrow [(\text{BBL}_{s1},\text{BBL}_{d1}), \dots, (\text{BBL}_{sn},\text{BBL}_{dn})]
$$

The key refers to a triplet, which contains two \emph{checkpoints} (\ie $cp_A$ and $cp_B$) and the hash of the \emph{LoA} (\ie \emph{H(LoA)}) associated to the significant BBLs that are traversed when moving from the source \emph{checkpoint} to the destination one. The value refers only to a subset of the BBLs pairs used to generate the hash of the \emph{LoAs} and, in particular, only to procedure calls and procedure returns. Those are the control-flow events required to mount the shadow stack during the verification phase.


\textbf{(3) Request for a Challenge.}  
The \emph{Verifier} starts a challenge with the \emph{Prover} by sending it an input and a nonce, which prevents replay attacks. 

\textbf{(4) Online Measurements.}
While the \emph{Application A} processes the input received from the \emph{Verifier}, the \emph{Prover} starts generating the \emph{online measurements} which keep trace of the \emph{Application A} executed paths. Each \emph{online measurement} is represented through the same notation used for the keys in the \emph{offline measurements}, \ie the triplet  $(\text{cp}_A,\text{cp}_B,H(\text{LoA}))$.

When the number of \emph{online measurements} reaches a preconfigured limit, 
the \emph{Prover} encloses all of them in a partial report and sends it to the \emph{Verifier}. The partial report is defined as follows: 
\begin{equation*}
	\begin{split}
	P_i &= (R,F_K(R||N||i))\\
	R &= (T, M).
	\end{split}
\end{equation*}
In the current notation, $P_i$ is the i-th partial report, $R$ the payload and $F_K(R||N||i)$ the digital fingerprint (\eg a message authentication code~\cite{bellare2000security}).
This is generated by using:
\begin{enumerate*}[label=(\roman*)]
	\item the secret key $K$, shared between \emph{Prover} and \emph{Verifier},
	\item the nonce $N$, sent at the beginning of the protocol, and
	\item the index $i$, which is a counter of the number of partial reports.
\end{enumerate*}
Finally, the payload $R$ contains the \emph{online measurements} $M$ along with the associated thread $T$.

The novel communication paradigm between \emph{Prover} and \emph{Verifier}, based on the transmission and consequent verification of several partial reports, satisfies the first requirement for applying runtime RA on complex systems (\ie handling the representation of a complex CFG/execution path). This is achieved thanks to the ScaRR control-flow model, which allows to fragment the whole CFG/execution path into sub-paths. Consequently, the \emph{Prover} can send intermediate reports even before the \emph{Application A} finishes to process the received input. In addition, the fragmentation of the whole execution path into sub-paths allows to have a more fine-grained analysis of the program runtime behaviour since it is possible to identify the specific edge on which the attack has been performed. 




\textbf{(5) Report Verification.}  
In runtime RA, the \emph{Verifier} has two different purposes: verifying whether the running application is still the original one and whether the execution paths traversed by it are the expected ones. The first purpose, which we assume to be already implemented in the system~\cite{costan2016intel,winter2008trusted}, can be achieved through a static RA applied on the \emph{Prover} software stack. On the contrary, the second purpose is the main focus in our design of the ScaRR schema. 

As soon as the \emph{Verifier} receives a partial report $P_i$, it first performs a formal integrity check by considering its fingerprint $F_K(R||N||i)$. Then, it considers the \emph{online measurements} sent within the report and performs the following checks: 
\begin{enumerate*}[label=(C\arabic*)]
    \item whether the \emph{online measurements} are the expected ones (\ie it compares the received \emph{online measurements} with the offline ones stored in the \emph{Measurements DB}),
    \item whether the destination \emph{checkpoint} of each \emph{measurement} is equal to the source \emph{checkpoint} of the following one, and
    \item whether the \emph{LoAs} are coherent with the stack status by mounting a shadow stack.
\end{enumerate*}
If one of the previous checks fails, the \emph{Verifier} notifies an anomaly and it will reject the output generated by the \emph{Prover}.

All the above-mentioned checks performed by the \emph{Verifier} are lightweight procedures (\ie a lookup in a hash map data structure and a shadow stack update). The speed of the second verification mechanism depends on the number of procedure calls and procedure returns found for each \emph{measurement}. Thus, also the second requirement for applying runtime RA on complex systems is satisfied (\ie keeping a fast verification phase). Once again, this is a consequence of the ScaRR control-flow model since the fragmentation of the execution paths allows both \emph{Prover} and \emph{Verifier} to work on a small amount of data. Moreover, since the \emph{Verifier} immediately validates a report as soon as it receives a new one, it can also detect an attack even before the \emph{Application A} has completed the processing of the input. 

\subsection{Shadow Stack}
\begin{figure}
	\centering
	\includegraphics[width=0.48\textwidth]{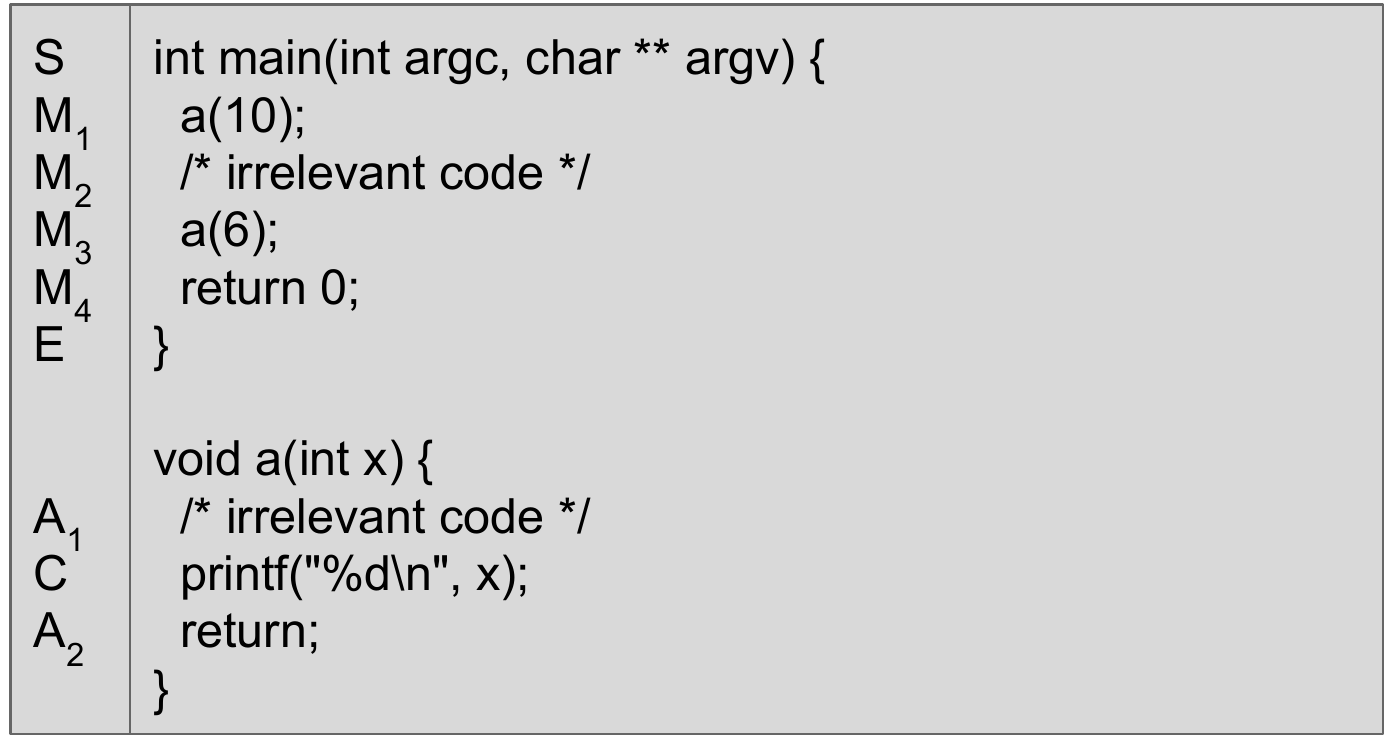}
	\caption{Illustrative example to explain the shadow stack on the ScaRR \emph{Verifier}.}
	\label{fig:trace-paths}
\end{figure}

To improve the defences provided by ScaRR, we introduce a shadow stack mechanism on the \emph{Verifier} side.
To illustrate it, we refer to the program shown in Figure~\ref{fig:trace-paths}, which contains only two functions:
\texttt{main()} and \texttt{a()}. Each line of the program is a BBL and, in particular: the first BBL (\ie \emph{S}) and the last BBL (\ie \emph{E}) of the \texttt{main()} function are a \emph{beginning thread} and \emph{end thread} \emph{checkpoints}, respectively; the function \texttt{a()} contains a function call to \texttt{printf()}, which is an \emph{exit-point}. 
According to the ScaRR control-flow model, the \emph{offline measurements} are the following ones:
\begin{align*}
(S,C,H_1) &\Rightarrow [(M_1,A_1)], \\
(C,C,H_2) &\Rightarrow [(A_2,M_2), (M_3,A_1)], \\
(C,E,H_3) &\Rightarrow [(A_2,M_4)].
\end{align*}
The significant BBLs we consider for generating the \emph{LoAs} are: \begin{enumerate*}[label=(\roman*)]
    \item the ones connecting the BBL S to the \emph{checkpoint} C,
    \item the ones connecting two \emph{checkpoints} C, and
    \item the ones to move from the \emph{checkpoint} C to the last BBL E.
\end{enumerate*}    

%

In this scenario, an attacker may hijack the return address of the function \texttt{a()} in order to jump to the BBL $M_3$.
If this happens, the \emph{Prover} produces the following \emph{online measurements}:
$$
(S,C,H_1) \rightarrow (C,C,H_2) \rightarrow (C,C,H_2) \rightarrow \dots.
$$
Although generated after an attack, those measurements are still compliant with the checks $(C1)$ and $(C2)$ of the \emph{Verifier}. Thus, to detect this attack, we introduce a new relation (\ie \texttt{ret\_to}) to illustrate the link between two edges. The \emph{Measurements Generator} computes all the \texttt{ret\_to} relations during the \emph{Offline Program Analysis} and saves them in the \emph{Measurements DB} using the following notation:
\begin{align*}
(A_2,M_2)~\texttt{ret\_to}~(M_1,A_1), \\
(A_2,M_4)~\texttt{ret\_to}~(M_3,A_1).
\end{align*}

\begin{figure}[t]
	\centering
	\includegraphics[width=0.4\textwidth]{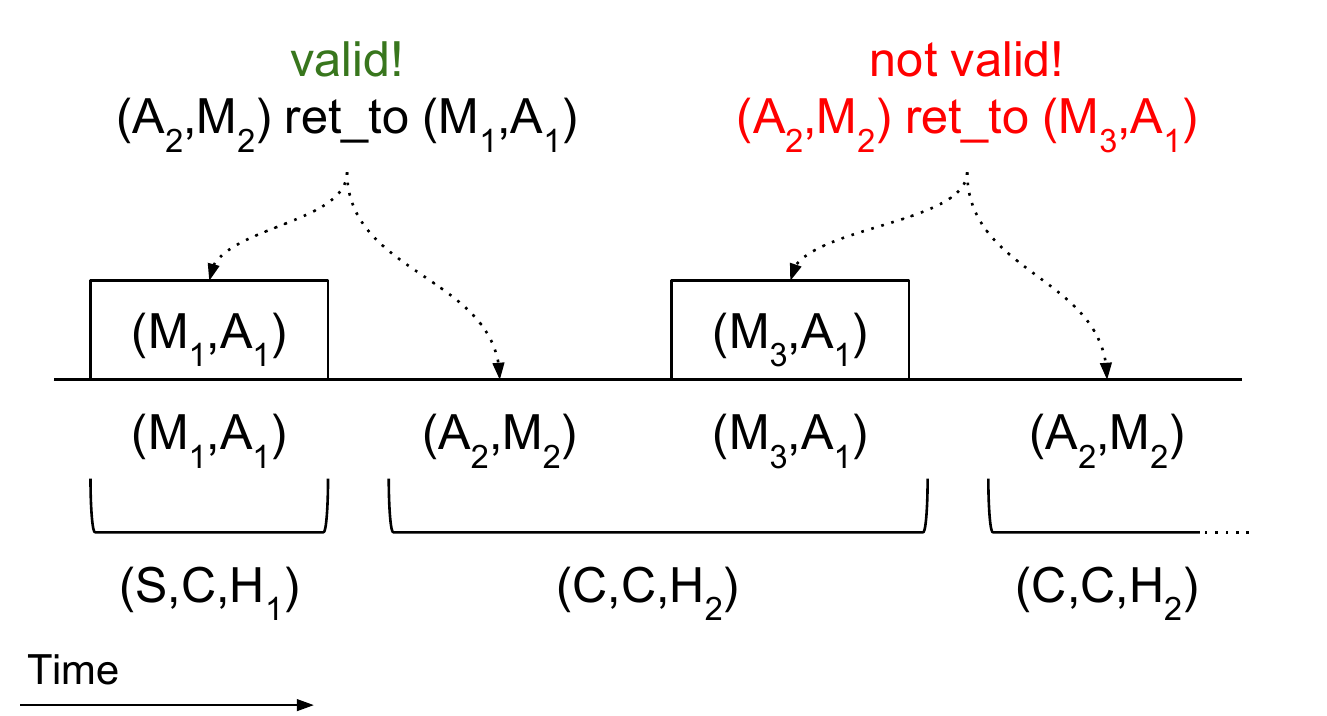}
	\caption{Implementation of the shadow stack on the ScaRR \emph{Verifier}.}
	\label{fig:shadow-stack}
\end{figure}

Figure~\ref{fig:shadow-stack} shows how the \emph{Verifier} combines all these information to build a remote shadow stack.
At the beginning, the shadow stack is empty (\ie no function has been invoked yet). Then, according to the \emph{online measurement} $(S,C,H_1)$, the \emph{Prover} has invoked the \texttt{main()} function passing through the edge $(M_1,A_1)$, which is pushed on the top of the stack by the \emph{Verifier}. Then, the \emph{online measurement} $(C,C,H_2)$ indicates that the execution path exited from the function \emph{a()} through the edge $(A_2,M_2)$, which is in relation with the edge on the  top of the stack and therefore is valid.
Moving forward, the \emph{Verifier} pops from the stack and pushes the edge $(M_3,A_1)$, which corresponds to the second invocation of the function \texttt{a()}.
At this point, the third measurement $(C,C,H_2)$ indicates that the \emph{Prover} exited from the function \texttt{a()}
through the edge $(A_2,M_2)$, which is not in relation with $(M_3,A_1)$. Thus, the  \emph{Verifier} detects the attack and triggers an alarm. 

\section{Implementation}
\label{sec:implementation}

Here, we provide the technical details of the ScaRR schema and, in particular, of the \emph{Measurements Generator} (Section~\ref{ssec:measurements_generator}) and of the \emph{Prover} (Section~\ref{ssec:prover}). 


\subsection{Measurements Generator}
\label{ssec:measurements_generator}
The \emph{Measurements Generator} is implemented as a compiler, based on LLVM~\cite{lattner2004llvm} and on the CRAB framework~\cite{gange2016abstract}. Despite this approach, it is also possible to use frameworks to lift the binary code to LLVM intermediate-representation (IR)~\cite{mcsema}.

The \emph{Measurements Generator} requires the program source code to perform the following operations:
\begin{enumerate*}[label=(\roman*)]
	\item generating the \emph{offline measurements}, and 
	\item detecting and instrumenting the control-flow events.
\end{enumerate*}
During the compilation, the \emph{Measurements Generator} analyzes the LLVM IR to identify the control-flow events and generate the \emph{offline measurements}, while it uses the CRAB LLVM framework to generate the CFG, since it provides a heap abstract domain that resolves indirect forward jumps.
Again during the compilation, the \emph{Measurements Generator} instruments each control-flow event to invoke a tracing function 
which is contained in the trusted anchor.
To map LLVM IR BBLs to assembly BBLs, we remove the optimization flags and we include dummy code,
which is removed after the compilation through a binary-rewriting tool.
To provide the above-mentioned functionalities, we add around $3.5$K lines of code on top of CRAB and LLVM $5.0$.


\subsection{Prover}
\label{ssec:prover}
\begin{figure}[t]
	\centering
	\includegraphics[width=0.5\textwidth]{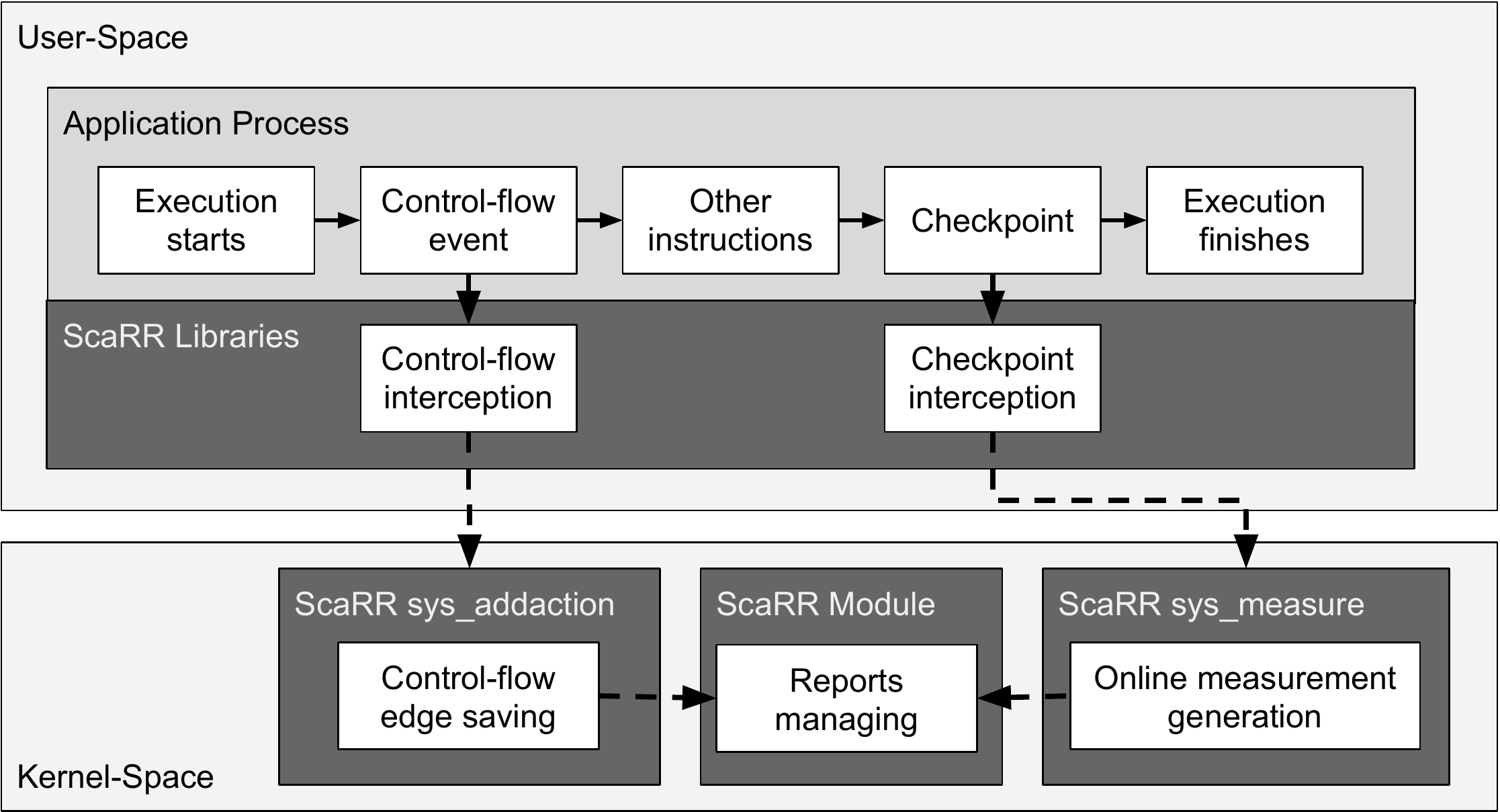}
	\caption{Internal architecture of the \emph{Prover}.}
	\label{fig:architecture}
\end{figure}

The \emph{Prover} is responsible for running the monitored application, generating the application \emph{online measurements} and sending the partial reports to the \emph{Verifier}. 
To achieve the second aim, the \emph{Prover} relies on the architecture depicted in Figure~\ref{fig:architecture}, which encompasses several components belonging either to the user-space (\ie \emph{Application Process} and \emph{ScaRR Libraries}) or to the kernel-space (\ie \emph{ScaRR sys\_addaction}, \emph{ScaRR Module}, and \emph{ScaRR sys\_measure}). 

Each component works as follows:  
\begin{itemize}
    \item \emph{Application Process} - the process running the monitored application, which is equipped with the required instrumentation for detecting control-flow events at runtime.
    \item \emph{ScaRR Libraries} - the libraries added to the original application to trace control-flow events and \emph{checkpoints}.
    \item \emph{ScaRR sys\_addaction} - a custom kernel syscall used to trace control-flow events.
    \item \emph{ScaRR Module} - a module that keeps trace of the \emph{online measurements} and of the partial reports. It also extracts the BBL labels from their runtime addresses, since the ASLR protection changes the BBLs location at each run.
    \item \emph{ScaRR sys\_measure} - a custom kernel syscall used to generate the \emph{online measurements}. 
\end{itemize}
When the \emph{Prover} receives a challenge, it starts the execution of the application and creates a new \emph{online measurement}.
During the execution, the application can encounter \emph{checkpoints} or control-flow events, both hooked by the instrumentation.
Every time the application crosses a control-flow event, the \emph{ScaRR Libraries}
invoke the \emph{ScaRR sys\_addaction} syscall to save the new edge in a buffer inside the kernel-space.
While, every time the application crosses a \emph{checkpoint}, the \emph{ScaRR Libraries}
invoke the \emph{ScaRR sys\_measure} syscall to save the \emph{checkpoint}
in the current \emph{online measurement}, calculate the hash of the edges saved so far, and,
finally, store the \emph{online measurement} in a buffer located in the kernel-space.
When the predefined number of \emph{online measurements} is reached, 
the \emph{Prover} sends a partial report to the \emph{Verifier} and starts collecting new \emph{online measurements}.
The \emph{Prover} sends the partial report by using a dedicated kernel thread.
The whole procedure is repeated until the application finishes processing the input of the \emph{Verifier}. 

The whole architecture of the \emph{Prover} relies on the kernel as a trusted anchor, since we find it more efficient in comparison to other commercial trusted platforms, such as SGX and TrustZone, but other approaches can be also considered (Section~\ref{sec:discussion}). To develop the kernel side of the architecture, we add around $200$ lines of code to a Kernel version v$4.17$-rc$3$.
We also include the Blake2 source~\cite{Aumasson2014,blake2}, which is faster and provides high cryptographic security guarantees for calculating the hash of the \emph{LoAs}.

\section{Evaluation}
\label{sec:evaluation}
We evaluate ScaRR from two perspectives.
First, we measure its performance focusing on: attestation speed (Section~\ref{ssec:attestation-speed}), verification speed (Section~\ref{ssec:verification-speed}) and network impact (Section~\ref{ssec:network-impact}).
Then, we discuss ScaRR security and privacy guarantees (Section~\ref{ssec:security+privacy+consideration}). 

We obtained the results described in this section by running the bench-marking suite SPEC CPU 2017 over a Linux machine
equipped with an Intel i7 processor and 16GB of memory~\footnote{We did not manage to map assembly BBL addresses to LLVM IR for 519.lbm\_r and 520.omnetpp\_r.}.
We instrumented each tool to detect all the necessary control-flow events, we then extracted the \emph{offline measurements} and we ran each experiment to analyze a specific performance metrics. 

\subsection{Attestation Speed}
\label{ssec:attestation-speed}
\begin{figure}[t]
	\centering
	\includegraphics[width=0.5\textwidth]{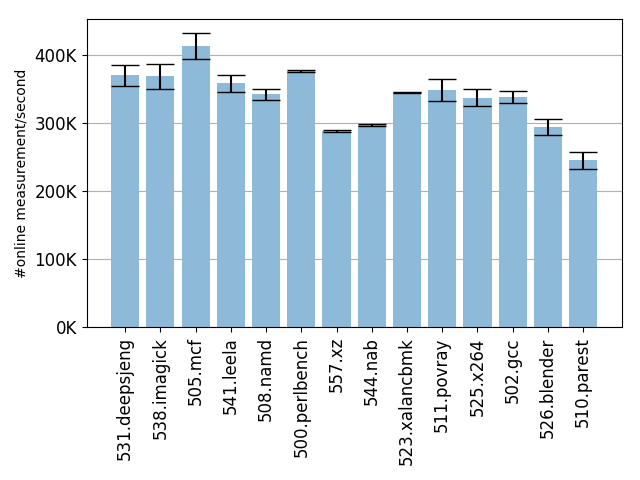}
	\caption{Average attestation speed measured as number of online measurements per second.}
	\label{fig:attesetation_speed}
\end{figure}

We measure the attestation speed as the number of \emph{online measurements} per second generated by the \emph{Prover}. 
Figure~\ref{fig:attesetation_speed} shows the average attestation speed and the standard deviation for each experiment of the SPEC CPU 2017. 
More specifically, we run each experiment 10 times, calculate the number of \emph{online measurements} generated per second in each run, and we compute the final average and standard deviation. 
Our results show that ScaRR has a range of attestation speed which goes from $250$K (510.parest) to over $400$K (505.mcf) of \emph{online measurements} per second. This variability in performance depends on the complexity of the single experiment and on other issues, such as the file loading. Previous works prove to have an attestation speed around $20K$/ $30K$ of control-flow events per second~\cite{aberadiat,abera2016c}. Since each \emph{online measurement} contains at least a control-flow event, we can claim that ScaRR has an attestation speed at least $10$ times faster than the one offered by the existing solutions.

\subsection{Verification Speed}
\label{ssec:verification-speed}
During the validation of the partial reports, the \emph{Verifier} performs a lookup against the \emph{Measurements DB} and an update of the shadow stack. 
To evaluate the overall performance of the \emph{Verifier}, we consider the verification speed as the maximum number of \emph{online measurements} verified per second. 
To measure this metrics, we perform the following experiment for each SPEC tool:
first, we use the \emph{Prover} to generate and save the \emph{online measurements} of a SPEC tool; 
then, the \emph{Verifier} verifies all of them without involving any element that might introduce delay (\eg network). 
In addition, we also introduce a digital fingerprint based on AES~\cite{Stallings:2002:AES:763194.763196} to simulate an ideal scenario in which the \emph{Prover} is fast. 
We perform the verification by loading the \emph{offline measurements} in an in-memory hash map and performing the shadow stack.
Finally, we compute the average verification speed of all tools.

According to our experiments, the average verification speed is 2M of \emph{online measurements} per second, with a range that goes from $1.4$M to $2.7$M of \emph{online measurements} per second. This result outperforms previous works in which the authors reported a verification speed that goes from $110$~\cite{Dessouky:2018:LLH:3240765.3240821} to $30$K~\cite{aberadiat} of control-flow events per second. As for the attestation speed, we recall that each \emph{online measurement} contains at least one control-flow event.

\begin{figure}[t]
	\centering
	\includegraphics[width=0.5\textwidth]{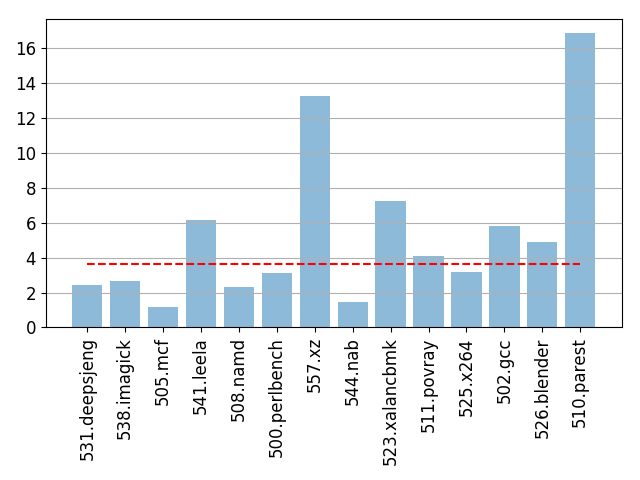}
	\caption{Average number of procedure calls and procedure returns found during the \emph{Online Program Analysis} of the SPEC CPU 2017 tools.}
	\label{fig:functioncall}
\end{figure}

The performance of the shadow stack depends on the number of 
procedure calls and procedure returns found during the generation of \emph{online measurements} in the \emph{Online Program Analysis} phase. 
To estimate the impact on the shadow stack, we run each experiment of the SPEC CPU 2017 tool and count the number of procedure calls and procedure returns. Figure~\ref{fig:functioncall} shows the average number of the above-mentioned variables found for each experiment. 
For some experiments (\ie 505.mcf and 544.nab), the average number is almost one since they include some recursive algorithms that correspond to small \emph{LoAs}. If the average length of the \emph{LoA}s tends to one, ScaRR behaves similarly to other remote RA solutions that are based on cumulative hashes~\cite{abera2016c,aberadiat}. Overall, Figure~\ref{fig:functioncall} shows that a median of push/pop operations is less than $4$, which implies a fast update.
Combining an in-memory hash map and a shadow stack allows ScaRR to perform a fast verification phase.

\subsection{Network Impact and Mitigation}
\label{ssec:network-impact}
\begin{figure}[t]
	\centering
	\includegraphics[width=0.5\textwidth]{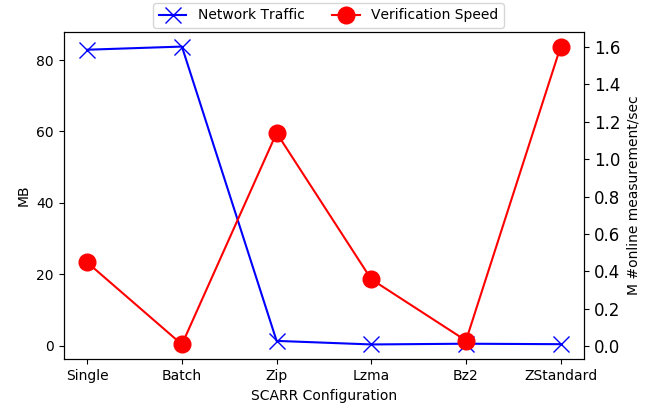}
	\caption{Comparison of different approaches for generating partial repors in terms of network traffic and verification speed.}
	\label{fig:network_performance}
\end{figure}

A high sending rate of partial reports from the \emph{Prover}
might generate a network congestion and therefore affect the verification phase.
To reduce network congestion and improve verification speed, we perform an empirical measurement of 
the amount of data (\ie MB) sent on a local network with respect to the verification speed by applying different settings.
The experiment setup is similar to Section~\ref{ssec:verification-speed}, but 
the \emph{Prover} and the \emph{Verifier} are connected through an Ethernet network with a bandwidth of $10$Mbit/s.
At first, we record $1$M of \emph{online measurements} for each SPEC CPU 2017 tool.
Then, we send the partial reports to the \emph{Verifier} over a TCP connection,
each time adopting a different approach among the following ones:
\emph{Single}, \emph{Batch}, \emph{Zip}~\cite{zip}, \emph{Lzma}~\cite{lzma}, \emph{Bz2}~\cite{bz2} and \emph{ZStandard}~\cite{zstandard}. 
The results of this experiment are shown in Figure~\ref{fig:network_performance}. 
In the first two modes (\ie \emph{Single} and  \emph{Batch}),
we send a single \emph{online measurement} and $50$K
\emph{online measurements} in each partial report, respectively.
As shown in the graph, both approaches generate a high amount of network traffic (around $80$MB),
introducing a network delay which slows down the verification speed.
For the other four approaches, each partial report still contains $50$K \emph{online measurements},
but it is generated through different compression algorithms.
All the four algorithms provide a high compression rate (on average over $95$\%) with a consequent reduction in the network overload.
However, the algorithms have also different compression and decompression delays, which affect the verification speed.
The \emph{Zip} and \emph{ZStandard} show the best performances with $1.2$M of \emph{online measurements}/s and $1.6$M of \emph{online measurements}/s, respectively, while \emph{Bz2} ($30$K of online measurements/s) and \emph{Lzma} ($0.4$M of online measurements/s) are the worst ones. 
The number of \emph{online measurements} per partial report might introduce a delay in
detecting attacks and its value depends on the monitored application.
We opted for $50$K because the SPEC CPU tools generate a high number of \emph{online measurements} overall.
However, this parameter strictly depends on the monitored application.
This experiment shows that we can use compression algorithms to mitigate the network congestion and keep a high verification speed.

\subsection{Security \& Privacy Considerations}
\label{ssec:security+privacy+consideration}

Here, we describe the security and privacy guarantees introduced by ScaRR. 

\textbf{Code Injection.}
In this scenario, an attacker loads malicious code, \eg \emph{Shellcode}, into memory and executes it by exploiting a memory corruption error~\cite{smith1997stack}.
A typical approach is to inject code into a buffer which is under the attacker control. 
The adversary can, then, exploit vulnerabilities (\eg buffer overflows) to hijack the program control-flow towards the shellcode (\eg by corrupting a function return address).


When a W$\oplus$X protection is in place, this attempt will generate a memory protection error, since the injected code is placed in a writable memory area and it is not executable. In case there is no W$\oplus$X enabled, the attack will generate a wrong \emph{LoA} detected by the \emph{Verifier}.

Another strategy might be to overwrite a node (\ie a BBL) already present in memory. 
Even though this attempt is mitigated by W$\oplus$X, as executable memory regions are not writable, it is still possible to perform the attack by changing the memory protection attributes through the operating system interface (\eg the \texttt{mprotect} system call in Linux), which makes the memory area writable. 
The final result would be an override of the application code. Thus, the static RA of ScaRR can spot the attack.


\textbf{Return-oriented Programming.}
Compared to previous attacks, the code-reuse ones are more challenging since they do not inject new nodes, but they simply reorder legitimate BBLs. Among those, the most popular attack~\cite{shacham2007geometry} is ROP~\cite{carlini2014rop}, which
exploits small sequences of code (gadgets) that end with a \texttt{ret} instruction. Those gadgets already exist in the programs or libraries code, therefore, no code is injected. The ROP attacks are Turing-complete in nontrivial programs~\cite{carlini2014rop}, and common defence mechanisms are still not strong enough to definitely stop this threat.

To perform a ROP attack, an adversary has to link together a set of gadgets through the so-called ROP chain, which is a list of gadget addresses. A ROP chain is typically injected through a stack overflow vulnerability, by writing the chain so that the first gadget address overlaps a function return address. Once the function returns, the ROP chain will be triggered and will execute the gadget in sequence. Through more advanced techniques such as stack pivoting~\cite{PracticalROP}, ROP can also be applied to other classes of vulnerabilities, \eg heap corruption.
Intuitively, a ROP attack produces a lot of new edges to concatenate all the gadgets, which means invalid \emph{online measurements} that will be detected by ScaRR at the first \emph{checkpoint}. 


\textbf{Jump-oriented Programming.}
An alternative to ROP attacks are the JOP ones~\cite{yao2013jop,bletsch2011jump}, which exploit special gadgets based on indirect \texttt{jump} and \texttt{call} instructions.
ScaRR can detect those attacks since they deviate from the original control-flow.

\textbf{Function Reuse Attacks.}
Those attacks rely on a sequence of subroutines, that are called in an unexpected order, \eg through virtual functions calls in C++ objects. ScaRR can detect these attacks, since the ScaRR control-flow model considers both the calling and the target addresses for each procedure call. Thus, an unexpected invocation will result in a wrong \emph{LoA}.
For instance, in Counterfeit Object-Oriented Programming (COOP) attacks~\cite{schuster2015counterfeit}, an attacker uses a loop to invoke a set of functions by overwriting a \emph{vtable} and invoking functions from different calling addresses generates unexpected \emph{LoAs}.

\section{Discussion}
\label{sec:discussion}
In this section we discuss limitations and possible solutions for ScaRR.

\textbf{Control-flow graph.}
Extracting a complete and correct CFG through static analysis is challenging.
While using CRAB as abstract domain framework, we experienced some problems
to infer the correct forward destinations in case of virtual functions. Thus, we will investigate new techniques to mitigate this limitation.

\textbf{Reducing context-switch overhead.}
ScaRR relies on a continuous context-switch between user-space and kernel-space.
As a first attempt, we evaluated SGX as a trusted platform, but we found out that the overhead was even higher due to SGX clearing the Translation-Lookaside Buffer~\cite{stravers2013translation} (TLB) at each enclave exit.
This caused frequent page walks after each enclave call.
A similar problem was related to the Page-Table Isolation~\cite{watson2018capability} (PTI) mechanism in the Linux kernel, which protects against the Meltdown vulnerability. 
With PTI enabled, TLB is partially flushed at every context switch, significantly increasing the overhead of syscalls.
New trusted platforms have been designed to overcome this problem, but, since they mainly address embedded software, they are not suitable for our purpose.
We also investigated technologies such as Intel PT~\cite{Ge:2017:GGC:3037697.3037716} to trace 
control-flow events at hardware level, but this would have bound ScaRR to a specific proprietary technology and we also found that previous works~\cite{Ge:2017:GGC:3037697.3037716,Hu:2018:EUC:3243734.3243797} experienced information loss.

\textbf{Physical attacks.}
Physical attacks are aimed at diverting normal control-flow such that the program is compromised, but the computed measurements are still valid. Trusted computing and RA usually provide protection against physical attacks. In our work, we mainly focus on runtime exploitation, considering that ScaRR is designed for a deployment on virtual machines. Therefore, we assume to have an adversary performing an attack from a remote location or from the user-space and the hosts not being able to be physically compromised. As a future work, we will investigate new solutions to prevent physical attacks.

\textbf{Data-flow attestation.}
ScaRR is designed to perform runtime RA over a program CFG. Pure data-oriented attacks might force the program to execute valid, but undesired paths without injecting new edges. To improve our solution, we will investigate possible strategies to mitigate this type of attacks, considering the availability of recent tools able to automatically run this kind of exploit~\cite{hu2016data}. 


\textbf{Toward a full semantic RA.}
We will investigate new approaches to validate series of \emph{online measurements} by using runtime abstract interpretation~\cite{Ge:2017:GGC:3037697.3037716,Hu:2018:EUC:3243734.3243797,Liu:2018:RED:3243734.3243826}.

\section{Related Work}
\label{sec:related-works}
Runtime RA shares properties with classic CFI techniques. Thus, we discuss current state-of-the-art of both research areas. 

\textbf{Remote Attestation.}
Existing RA schemes are based on a cryptographic signature of a piece of software (\eg software modules, BIOS, operating system). Commercial solutions that implement such mechanisms are already available: TPM~\cite{tomlinson2017introduction}, SGX~\cite{costan2016intel}, and AMD TrustZone~\cite{winter2008trusted}. Academic approaches, which focus on cloud systems, are proposed by Liangmin et al.~\cite{wang2018trusted} and Haihe et al.~\cite{ba2017jmonatt}. More specifically, their solutions involve a static attestation schema for infrastructures as a service and JVM cloud computing, respectively. Even though these technologies can provide high-security guarantees, they focus on static properties (\ie signatures of components) and cannot offer any defence against runtime attacks. 


To overcome design limitations of static RA, researchers propose
runtime RA. Kil et al.~\cite{kil2009remote} analyze base pointers of software components, such as stack and heap, and compare them with the measurements acquired offline. Bailey et al.~\cite{bailey2010trusted} propose a coarse-grained level that attests the order in which applications modules are executed. Davi et al.~\cite{davi2009dynamic} propose a runtime attestation based on policies, such as the number of instructions executed between two consecutive returns. Previous works suggest first to acquire a runtime measurement of software properties, but do not provide a fine-grained control-flow analysis.

A modern fine-grained control-flow RA is represented by C-FLAT, which is proposed by Abera et al.~\cite{abera2016c}. This schema measures the valid execution paths undertaken by embedded systems and generates a hash, which length depends on the number of control-flow events encountered at runtime. Then, the hash is compared with a list of offline measurements. 
The main differences between ScaRR and C-FLAT are the following ones: 
\begin{enumerate*}[label=(\roman*)]
	\item C-FLAT control-flow representation grows along with software complexity, while ScaRR  manages complex control-flow paths by using partial reports, and
	\item ScaRR is designed to use features of modern computer architectures (\eg multi-threading, bigger buffers).
\end{enumerate*}
Dessouky et al. propose LO-FAT~\cite{dessouky2017fat},
which is a C-FLAT hardware implementation aimed at improving runtime performances for embedded systems. However, LO-FAT inherits all of C-FLAT design limitations in term of control-flow representation. Zeitouni et al. designed ATRIUM~\cite{zeitouni2017atrium}, that strengthens runtime RA schemes against physical attacks for embedded devices. Even though the authors address different use cases, this solution might be combined with ScaRR. 

Dessouky et al. propose LiteHax~\cite{Dessouky:2018:LLH:3240765.3240821}, that deals with data-only attacks. Their approach shares some similarities with ScaRR: they send detailed control-flow events information to a \emph{Verifier}. However, they target data-oriented attacks (instead of control-flow). Moreover, LiteHax uses symbolic execution to validate the reports, which slows down the verification phase. Abera et al. discuss DIAT~\cite{aberadiat}, which is a scalable RA for collaborative autonomous system. They model a runtime control-flow as a multi-set. This allows DIAT to represent complex control-flow graphs by using a relatively short hash. However, its model loses information about the execution order of the branches. This makes their approach prone to attacks like COOP~\cite{schuster2015counterfeit}. ScaRR, instead, combines a strong static analysis and a shadow execution at the \emph{Verifier} side that provides a sound approach by design. Overall, our experiments show that ScaRR can handle a higher number of branches per second compared to all the state-of-the-art runtime RA schemes.

\textbf{Control-Flow Integrity.}
In the last few years, some authors have proposed architectures that share some similarities with RA~\cite{Ding:2017:EPP:3241189.3241201,Liu:2018:RED:3243734.3243826,Hu:2018:EUC:3243734.3243797}. These works are composed by two concurrent processes: a target process (that might be under attack), and a monitor process (that validate some target property).
However, ScaRR considers a different attacker model since we consider a fully compromised user-space, \ie an attacker may tamper with the target software code or attack the monitor process itself. 
Moreover, unlike ScaRR, these solutions are not designed to provide any report about the execution path of the target process. 



\section{Conclusion}
\label{sec:conclusion}


In this work, we propose ScaRR, the first schema that enables runtime RA for complex systems to detect control-flow attacks generated in user-space.
ScaRR relies on a novel control-flow model that allows to:
\begin{enumerate*}[label=(\roman*)]
	\item apply runtime RA on any software regardless of its complexity,
	\item have intermediate verification of the monitored program, and
	\item obtain a more fine-grained report of an incoming attack.
\end{enumerate*}

We developed ScaRR and evaluated its performance against the set of tools of the SPEC CPU 2017 suite.
As a result, ScaRR outperforms existing solutions for runtime RA on complex systems in terms of attestation and verification speed, while guaranteeing a limited network traffic.

Future works include: investigating techniques to extract more precise CFG, facing compromised operating systems, and 
studying new verification methods for partial reports.

\newpage

\bibliographystyle{plain}
\bibliography{biblio}

\end{document}